\documentclass{article}
\usepackage[T1]{fontenc} 
\usepackage[utf8]{inputenc} 
\usepackage{ismir,amsmath,cite,url}
\usepackage{graphicx}
\usepackage{color}

\usepackage[firstpage]{draftwatermark}
\definecolor{lightgray}{rgb}{0.9,0.9,0.9}
\definecolor{darkgray}{rgb}{0.4,0.4,0.4}
\SetWatermarkFontSize{12pt}
\SetWatermarkScale{1.1}
\SetWatermarkAngle{90}
\SetWatermarkHorCenter{202mm}
\SetWatermarkVerCenter{170mm}
\SetWatermarkColor{darkgray}
\SetWatermarkText{Late-Breaking / Demo Session Extended Abstract, ISMIR 2024 Conference}

\def\lbd{}	

\usepackage[bookmarks=false,pdfauthor={Daeun Hwang},pdfsubject={Paper Title Here},hidelinks]{hyperref}
\title{A Music Information Retrieval Approach to Classify Sub-Genres in Role Playing Games}

\multauthor
{Daeun Hwang$^1$ \hspace{1cm} Xuyuan Cai$^1$ \hspace{1cm} Edward F. Melcer$^1$} { \bfseries{Elin Carstensdottir$^1$}\\
 $^1$ Computational Media, University of California, Santa Cruz, USA\\
{\tt\small dhwang8@ucsc.edu, xcai30@ucsc.edu}}

\def\authorname{D. Hwang, X. Cai, E. Melcer, and E. Carstensdottir}

\begin{document}
\maketitle
\begin{abstract}
Video game music (VGM) is often studied under the same lens as film music, which largely focuses on its theoretical functionality with relation to the identified genres of the media. However, till date, we are unaware of any systematic approach that analyzes the quantifiable musical features in VGM across several identified game genres. Therefore, we extracted musical features from VGM in games from three sub-genres of Role-Playing Games (RPG), and then hypothesized how different musical features are correlated to the perceptions and portrayals of each genre. This observed correlation may be used to further suggest such features are relevant to the expected storytelling elements or play mechanics associated with the sub-genre.
\end{abstract}
\section{Introduction}\label{sec:introduction}
As video game media are continuously gaining popularity, video game music (VGM) has established itself as a unique genre in the music industry. VGM is often studied under the same lens as film music, which largely focuses on its theoretical functionality with relation to the identified genres of the media. Although there are many studies that offer a systematic quantitative approach in musical information retrieval for film music \cite{ma2021computational}, there is a gap for investigating VGM with relation to game genres. VGM is largely different from traditional film scores, as video games are fundamentally different from films as media forms \cite{jorgensen2011time, collins2013playing}. 

With that in mind, we investigated the musical features of soundtracks spanning from three identified video game subgenres: Adventure Role-Playing games, Action Role-Playing games, and Strategy Role-Playing games. Currently, there are mainly two aspects that game scholars look at when classifying game genres: game mechanics (how the game is played) and game narratives (what the game is about) \cite{wolf2001medium}. In fact, genre classification in any media is heavily involved with human perception. Although these two aspects are very important in game studies for which they offer a reliable and approachable perspective when doing game analysis, sound and music are often less considered in existing studies. Games in the early days that had minimal storage capacity still incorporated sound effects and even music \cite{collins2013playing}. By today’s standards, it is almost unthinkable for games to not have any sound and music accompanying the gameplay. It is evident that music is an integral aspect in video games. Applying feature extraction techniques on VGM, we can investigate how the compositions of VGM are affected by the identification of game genres, and further contribute to the studies of genre classification in game studies. 

\begin{figure}
 \centerline{\framebox{
 \includegraphics[alt={ISMIR 2024 LBD template test image},width=0.9\columnwidth]{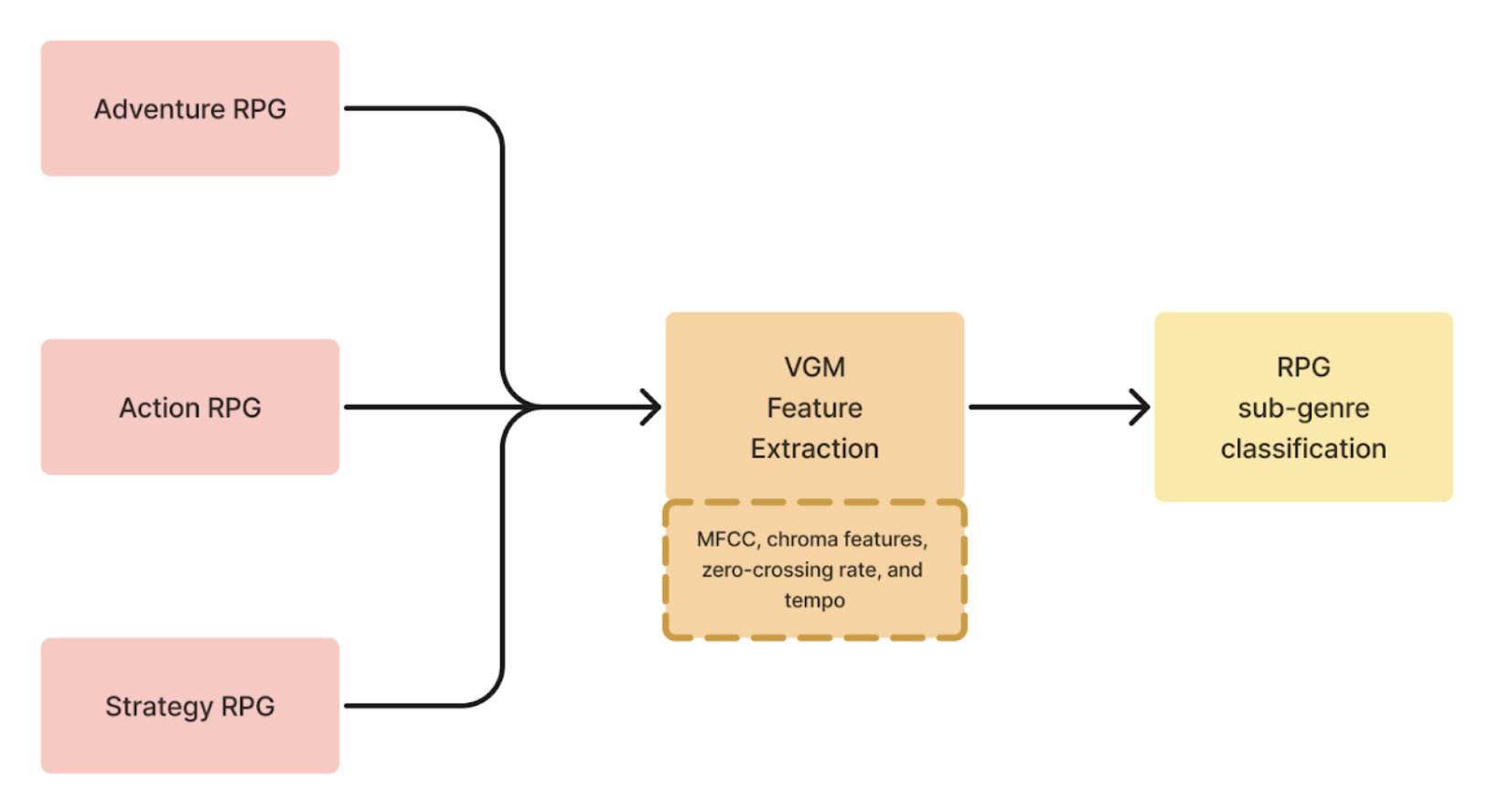}}}
 \caption{Overview of musical feature characterization for RPG sub-genres}
 \label{fig:overview}
\end{figure}

\section{BACKGROUND}
Over time, there has been an increasing focus on musical genre classification and feature extraction, which is a key method used to classify and characterize genres. McKay and Fujinaga highlighted the potential of using computing approaches for music genre classification \cite{mckay2006musical}. This approach was initially more prevalent in film music before being applied to game music \cite{ma2021computational}. 
When feature extraction is employed for musical genre classification, it first requires a dataset for analysis and training \cite{mckay2004automatic} The type of feature focused on to characterize music genres varies widely, but the most frequently used features are symbolic and audio features \cite{tzanetakis2003pitch}. Within symbolic and audio features, past studies have utilized pitch, scales, Mel Frequency Cepstral Coefficients (MFCC), among others \cite{mutiara2016musical, tzanetakis2003pitch, choudhury2013melodic}.
However, to the extent of our knowledge, there has not been any study that focused on music genre classification in game studies, which reveals that there has been a lack of research on game background music using computational approaches.

\section{METHOD}
In consideration of feasibility and accessibility, we used Steam\footnote{Web-based video game digital distribution service https://store.steampowered.com} as our platform to pick games under these three sub-genres based on popularity. The three most popular sub-genres under RPG were identified by Steam. After selecting top three most popular games under each sub-genres, we randomly selected three instrumental soundtracks from each game to construct a dataset. We avoided lyrical tracks because they are not necessarily apparent in all the games, while instrumental tracks are the more dominant type of music in game soundtracks. In total, we collected 27 instrumental soundtracks, which span from 9 different games that are classified under 3 different RPG sub-genres as suggested by Steam.
 
In order to analyze features of the songs, we first had to preprocess the dataset. First, all the files were normalized to -5 dB to match the decibel levels, and then were  trimmed into 15 seconds. This 15 seconds was in the middle of the song to approximately capture the highlighted part, or the main theme of the song. The sample rate was set to 48000 Hz to make the analysis consistent. Finally, all files were converted into WAV format in order to apply feature extractions on python.

The librosa library \cite{mcfee2015librosa} built-in method was mainly used to calculate and compute the following variables: average tempo in beats per minute (BPM), Zero-Crossing Rate (ZCR), average spectral centroid, average chroma feature, and MFCCs. The demonstration on Google Colab Notebook can be accessed through the link\footnote{Demo: {\href{https://colab.research.google.com/drive/1Fmwu6UKqbyrd-ih9WkgDxRu3ZDFBzvTm?usp=sharing}{https://colab.research.google.com/drive/1Fmwu6UKqbyrd-ih9WkgDxRu3ZDFBzvTm?usp=sharing}}}. 

\section{Results}
Overall pattern of average MFCCs in songs from three sub-genres seem to not have significantly distinct differences. Meanwhile, relatively the range of MFCC coefficients for adventure RPG seem to have the widest range, while strategy RPG has the narrowest range.

The average ZCR for Adventure RPGs seems to fluctuate more than the average ZCR for the other two sub-genres. This indicates that soundtracks in Adventure RPGs are probably more robust in terms of their audio signals, which hints the corresponding robustness of gameplay (for example, multiple thematic locations, multiple emotions, multiple game mechanics, etc.) in Adventure RPGs when compared to Strategy RPGs and Action RPGs.

For the average spectral centroid, the genre of RPG game did not seem to create a huge difference. This implies that the overall brightness of the sound \cite{grey1978perceptual} is quite in the similar level for music across sub-genres of RPG.

The average chroma features for Adventure RPGs seems to have a wider variety of key distribution for the melodies, where there are not as many specific keys that stood out when compared to the other two sub-genres. This further supports the pattern we inspected in the ZCR features previously where it also suggests that Adventure RPGs might be more robust than SRPGs and Action RPGs as they used a variety of musical keys in their melodies. 

The average tempo across all three sub-genres are fairly similar, there was no significant difference observed. One possible explanation for this is the fact that we are only using instrumental tracks, which might share similar tempo as opposed to lyrical tracks.

When we trained and tested a model using KNN model \cite{guo2003knn}, it demonstrated an accuracy of 33.3\%, which is not significant enough to be considered reliable. This suggests that further training using other algorithms or expanding the dataset may be necessary.

\section{LIMITATION AND FUTURE WORK}
Our study focused solely on RPGs and their sub-genre classifications. While this provides insights into the musical feature characterizations of music in RPGs, we plan to expand this research to include a broader range of games. Additionally, including more songs within this dataset would enhance the accuracy and concreteness of the analysis of musical feature characteristics of VGM in different game genres. 

\section{Conclusion}
Inspired by the MIR techniques from film music research, we approached VGM in hope to provide a different perspective when analyzing game music. Through musical feature extraction, we visualized multiple extraction methods and identified a few significant patterns. While this may not immediately provide a concrete relationship between musical features and the identified genre, we argue that the very existence of such observed correlation between musical features and the identified genre merits research potentials.

\bibliography{ISMIRtemplate}

\end{document}